\documentclass[sigconf]{acmart}
\usepackage{graphicx}
\usepackage{float} 
\usepackage{subfigure}
\usepackage{enumitem}
\usepackage{multirow}
\usepackage{makecell}
\usepackage{bm}
\usepackage{fixmath}

\AtBeginDocument{%
  \providecommand\BibTeX{{%
    \normalfont B\kern-0.5em{\scshape i\kern-0.25em b}\kern-0.8em\TeX}}}

\setcopyright{acmcopyright}
\copyrightyear{2020}
\acmYear{2020}
\acmDOI{10.1145/1122445.1122456}

\renewcommand\footnotetextcopyrightpermission[1]{} 
\acmPrice{15.00}
\acmISBN{978-1-4503-XXXX-X/18/06}

\author{Lifang Deng, Jin Niu, Angulia Yang, Qidi Xu, Xiang Fu, Jiandong Zhang, Anxiang Zeng}
\affiliation{%
  \institution{Alibaba Group, Beijing, China.}
  }
\begin{document}

\title{Hybrid Interest Modeling for Long-tailed Users}
\begin{abstract}
User behavior modeling is a key technique for recommender systems. However, most methods focus on head users with large-scale interactions and hence suffer from data sparsity issues. Several solutions integrate side information such as demographic features and product reviews, another is to transfer knowledge from other rich data sources. We argue that current methods are limited by the strict privacy policy and have low scalability in real-world applications and few works consider the behavioral characteristics behind long-tailed users. In this work, we propose the Hybrid Interest Modeling~(HIM) network to hybrid both personalized interest and semi-personalized interest in learning long-tailed users' preferences in the recommendation. To achieve this, we first design the User Behavior Pyramid~(UBP) module to capture the fine-grained personalized interest of high confidence from sparse even noisy positive feedbacks. 
Moreover, the individual interaction is too sparse and not enough for modeling user interest adequately, we design the User Behavior Clustering~(UBC) module to learn latent user interest groups with self-supervised learning mechanism novelly, which capture coarse-grained semi-personalized interest from group-item interaction data. Extensive experiments on both public and industrial datasets verify the superiority of HIM compared with the state-of-the-art baselines.


\end{abstract}


\maketitle

\section{Introduction}
Benefit from the widespread usage of the Internet and mobile technologies, personalized recommender systems live at the heart of the industry, which aim to provide customized items for each user. User behavior modeling is a key technique for recommender systems in which historical interactions play a crucial role~\cite{zhou2019deep,sun2019bert4rec,pi2019practice}.

\begin{figure}[t]
\subfigure[Taobao]{
\label{Fig.sub.tb}
\includegraphics[width=0.22\textwidth]{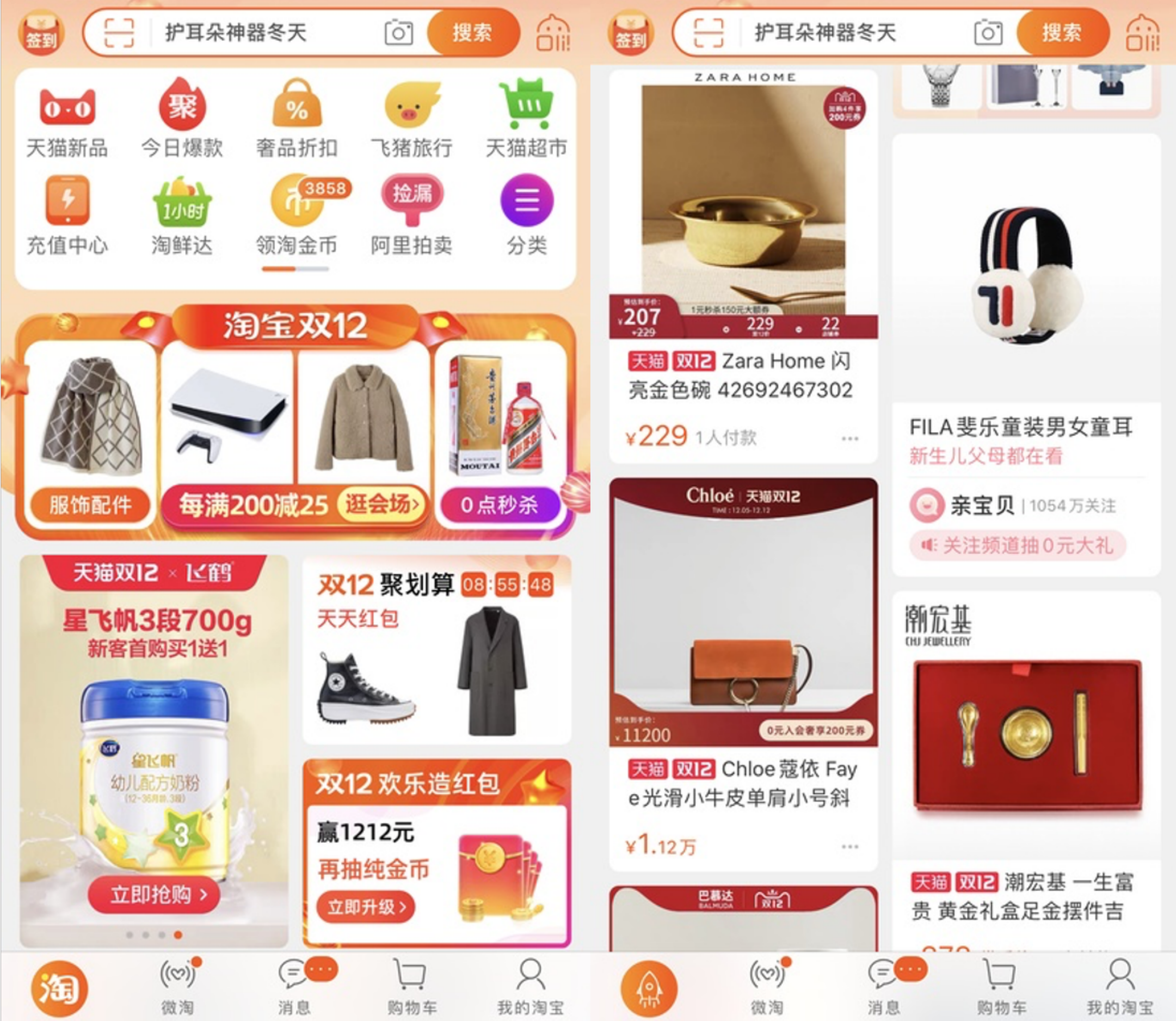}}
\subfigure[Lazada]{
\label{Fig.sub.lazada}
\includegraphics[width=0.22\textwidth]{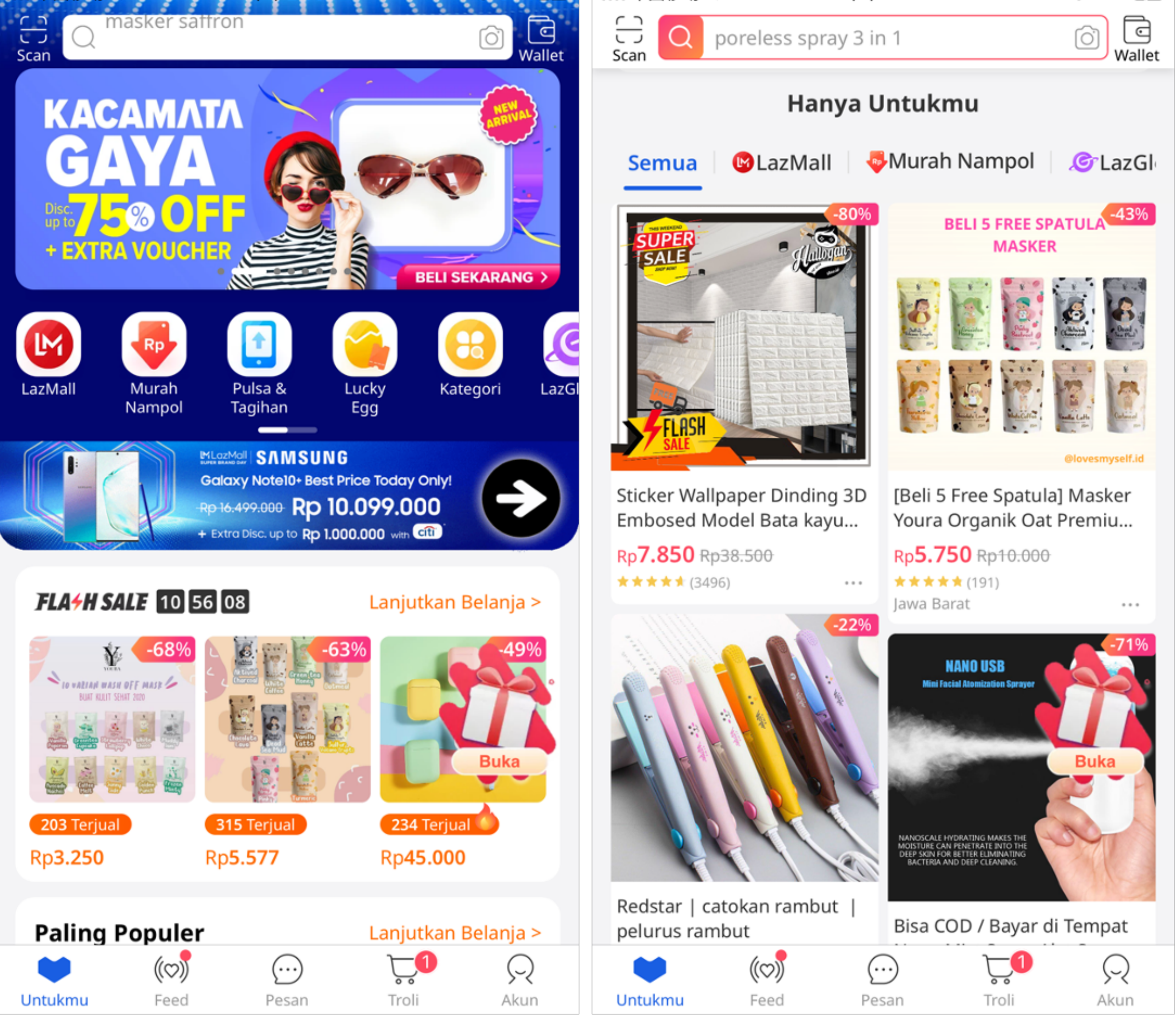}}
\caption{Homepage recommendation in Taobao and Lazada.}
\label{fig:rs_homepag}
\end{figure}

\begin{figure}[thtp]
\centering 
\includegraphics[width =5cm]{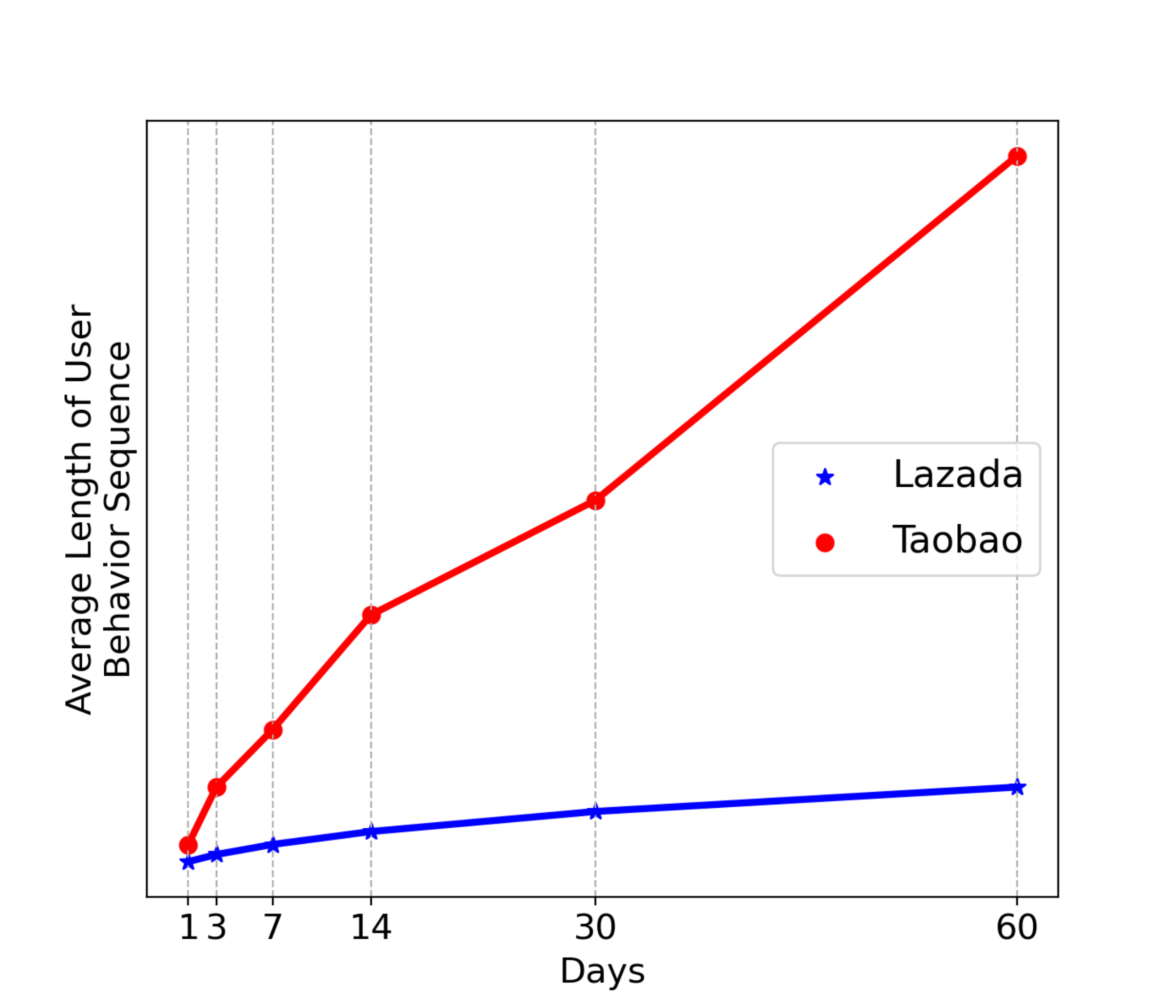}
\caption{Statistics of user interactions.}
\label{fig:user_data}
\end{figure}

Most methods design complex model to capture user’s latent interest, a wealth of information rest in their abundant user interactions. User-item rich interactions prop up the accurate recommendation and good recommendation greatly decreases user churn rate. Such a running mode creates a positive feedback loop. However, most users do not express their interests explicitly~(e.g., click, buy), the number of user’s interactions inherently follows a long-tailed distribution in real-world recommendation applications. Head users take their advantages in data richness to be well-served, while tail users are just the opposite. 
Such long-tailed users are silent majority in recommender systems, especially for those rising companies. Taking two recommender systems in the homepages of Taobao and Lazada for example~(illustrated in Figure~\ref{fig:rs_homepag}).
Lazada is a rising e-commerce company in Southeast Asia. The user scale and interactions numbers are much smaller than Taobao. 
The visualized data difference is presented in Figure~\ref{fig:user_data} (for privacy concern, we only visualize tendency and eliminate details), the average length of user behavior sequences in Taobao has a large increase than Lazada as time going. Long-tailed users dominate such rising e-commerce companies but are ill-served by the current user behavior modeling methods.

However, implementing high-performance recommendation for long-tailed users will also be full of challenges. 
Several solutions integrate side information such as demographic features and product reviews or to transfer knowledge from other rich data sources. We argue that current methods may have low scalability in real-world applications and raise privacy concerns.
Another promising way is group recommendation. such methods capture coarse-grained semi-personalized interest from group-item interaction where user preferences on an unobserved item can be related to the group they belong to~\cite{cao2018attentive,ma2019dbrec}. 
But these methods either do not make use of a single user's individual interactions to capture fine-grained personalized interest or need extra group information for initialization, making these methods hard to be deployed online.

In this paper, we propose a Hybrid Interest Modeling~(HIM) network to improve the recommendation performance of long-tailed users in an end-to-end manner without the requirement of auxiliary data sources. To achieve this, we first take full consideration of behavior patterns behind long-tailed users. These users are a massive quick-changing group with few interactions and large intervals.
We divide the long time interval interactions into multiple sessions and capture the users' personalized interest and semi-personalized interest within each period. We first design the User Behavior Pyramid module~(UBP) to capture fine-grained personalized interest. In this module, we consider user may click an item by mistake with no actual interest. This can bring a worse experience for long-tailed users than for head users due to fewer interactions. To conquer this, UBP takes both the positive(e.g., click) and negative(e.g., not click) interactions as inputs and models the differences between them to selectively extract intrinsic interest of high confidence from sparse even noisy positive interactions. The key idea is that interactions with higher confidence contribute more to the accurate recommendation. Moreover,  individual interaction is too sparse and not enough for modeling user interest adequately. We further design the User Behavior Clustering~(UBC) module to learn latent user interest groups with self-supervised learning mechanism novelly in an end to end manner, which captures coarse-grained semi-personalized interest from group-item interaction data. HIM models user-item interactions as well as group-item interactions. The two learning processes reinforce each other and boost recommendation performance for long-tailed users.


To sum up, HIM contributes to the following aspects:
\begin{enumerate}[topsep=7pt]
\item We find that existing recommendation models ignore the behavior characteristics of long-tailed users. We explore long-tailed users' behavior patterns and reorganize interactions in an effective way.
\item We propose HIM to improve user behavior modeling for long-tailed users and two specially designed modules UBP and UBC can effectively reduce data sparsity and boost recommendation performance. 
\item We validate on both public and industrial datasets and verify the superiority of HIM compared with the state-of-the-art baselines. It is notable that HIM has been deployed in Lazada online recommender systems and has obtained more than 7\% item page view~(IPV) improvements across multiple Southeast Asian countries. 
\end{enumerate}

\section{Related Work}
\label{sec:relatedwork}
\subsection{Deep Learning for Recommendation}
Deep learning has led to many successes in building industry-scale recommender systems~\cite{cheng2016wide,covington2016deep,he2017neural}.
Compared to conventional approaches~\cite{koren2009matrix, beutel2017beyond,chen2016xgboost,he2014practical}, deep learning models capture underlying relationships between users and items more effectively. A line of research put their efforts on discovering higher-order interactions among features, additionally start to replace manual features crafting with deep feature extraction~\cite{liu2019feature,he2017neural,guo2017deepfm,qu2016product,shan2016deep}. 
These models focus on mining the static relationships between users and items, ignoring the dynamics of users’ preferences in real-world recommendation scenarios. 
Therefore several deep recommendation methods concentrate on capturing users' preference from the rich historical interactions~\cite{zhou2018deep, pi2019practice,chen2017entive, wsdm-ctr2,sun2019bert4rec}. 
However, these methods heavily depend on the scale and quality of historical interactions. 
They ignore the problem of long-tailed distribution, which may cause performance degradation for tail users with limited interactions. 

To address this issue, existing methods generally integrate auxiliary information of different modalities such as user profiles, products review, title, or other item side information to boost recommendation performance with various deep model architectures~\cite{zheng2017joint,he2017neural,fu2019deeply,bansal2016ask,hu2017integrating,hu2019transfer}. However, the accessibility of additional auxiliary information limits their scalability when they are deployed in different recommendation scenarios and raise privacy concerns.

\begin{figure*}[thbp]
  \vspace{-0.5cm}
  \centering
  \includegraphics[width=0.95\textwidth]{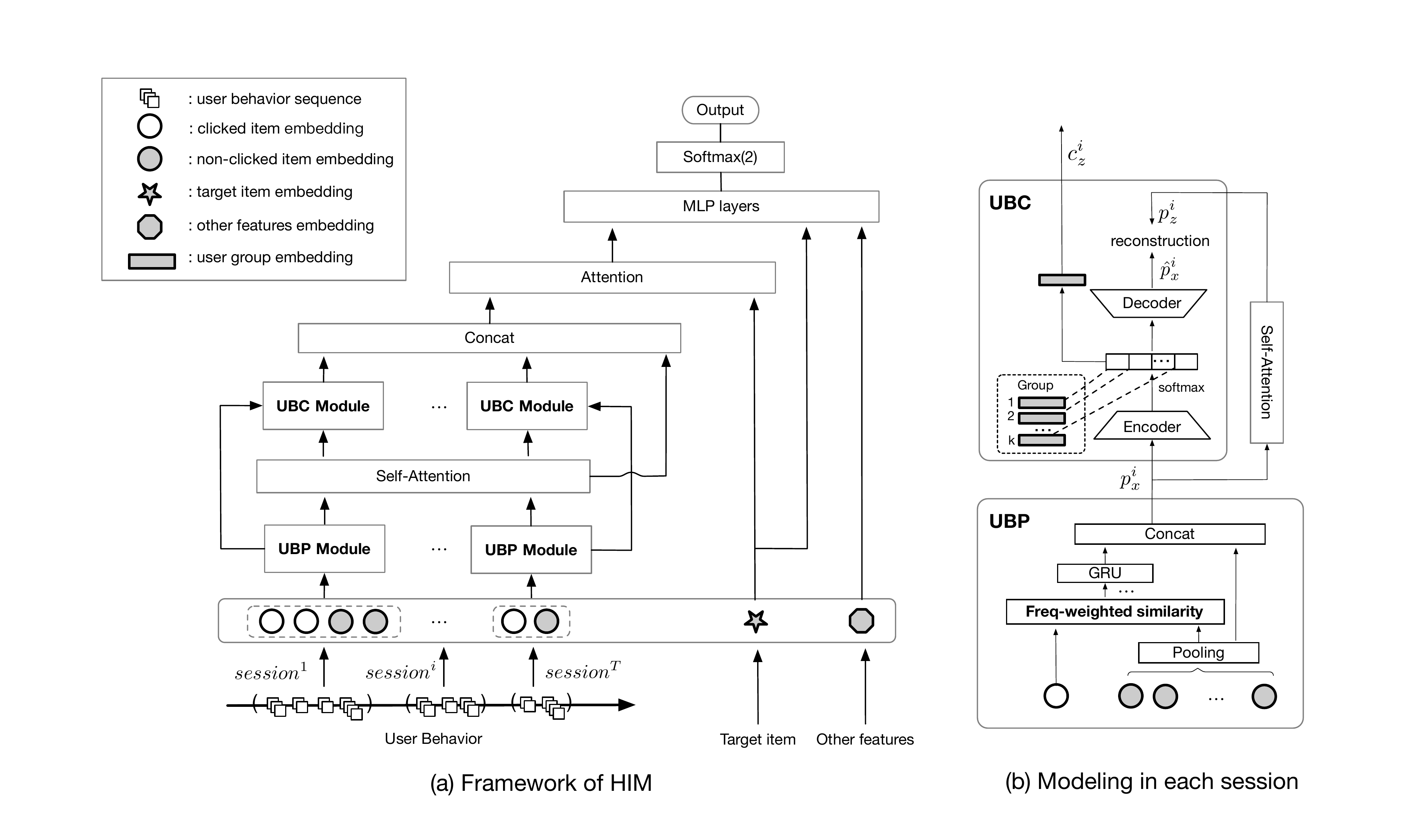}
  \caption{The framework of HIM and inner modules design. (a) The framework of HIM: At the reorganization layer, users' behavior sequences are split into different sessions and rearranged based on interaction frequency. HIM contains two modules UBP and UBC to model personalized interest and semi-personalized interest, respectively. The attention layer models the interests hybrid process that is relative to the target item and obtains compact attentive features. The weighted user interest representations and remaining features are concatenated and fed into shared MLP for the final prediction. (b) UBP and UBC design: In UBP, we take both the positive and negative feedback as inputs and model the relationship between them from the confidence perspective, we utilize GRU and self-attention units to obtain enhanced feature representations. In UBC, we apply Auto-encoder to learn latent user interest groups by constructing a self-supervised mechanism based on the UBP personalized representations.}
  \label{fig:approach}
\end{figure*}



\subsection{Group Recommendation}
Group recommendation received a lot of attention in recent years and has been widely applied in various domains. Most previous group recommendation methods rely on pre-defined users' group information such as social relation~\cite{cao2018attentive,vinh2019interact,seko2011group,hu2014deep} to give recommendation to a group of people. AGRRE~\cite{cao2018attentive} exploits attention mechanisms with group embeddings. 
DBRec~\cite{ma2019dbrec} uses a dual-bridge framework that initializes with pre-trained embeddings jointly integrate collaborative filtering, latent group discovery and hierarchical modeling tasks into a unified network. 
PreHash~\cite{shi2020beyond} defines anchor users and groups users into each bucket by a hashing network. However,
these methods need to be well initialized, the group information or the anchor user is explicitly required,
hence those methods cannot be scaled to the scenarios where the initialization information is not explicitly available.

\section{Hybrid Interest Modeling~(HIM)}
\label{sec:methods}
Based on the sparse interactions dilemma for long-tailed users and deficiency of previous methods, we proposed Hybrid Interest Modeling~(HIM) network. The model structure is shown in Figure~\ref{fig:approach}. In the following subsection, we will illustrate each module of HIM in detail.

\subsection{Reorganization Layer}
In the e-commerce recommender system, users can click items, add items to cart, or buy items. We treat these behaviors actively performed by users as positive feedback. If users only browse the recommended item with no further action, we mark it as negative feedback which is usually ignored in previous user behavior modeling. Here we argue that negative feedback can potentially indicate items they are not interested in and can be used as supplementary information to alleviate the sparsity problem of positive feedback. Thus we collect both positive feedback and negative feedback sequences as user behavior input to enrich user interest learning.

As for interactions reorganization, we observe that interactions of long-tailed users have larger time interval and informative temporal dependency is rare, the interactions often shows session-based traits. Thus following the idea of session-based recommendation~\cite{hidasi2015session}, we propose to divide the entire long behavior sequence into $T$ sessions. The specific length of
the session T is a hyperparameter that we can adjust to get the
optimal performance by user behavior distribution.  

Then, in each session, we collect one user's interactions and re-rank them by interaction frequency instead of temporal order. We argue that more frequent interactions have greater confidence in indicating user authentic interest. For example, if a user clicks a mobile phone 8 times, the user may have a greater preference for the mobile phone than other items clicked once. Both the positive and negative feedback  are reranked in frequently descending order. 

Besides, HIM adopts target item components and other related features as input as well, which is a commonly used pre-processing operation. We omit the details here.

\subsection{User Behavior Pyramid~(UBP)}
The above-mentioned Reorganization Layer is the basis for modeling long-tailed users' behavior characteristics. Based on the reorganization module, the User Behavior Pyramid~(UBP) is built to further capture users' interest within a session by modeling the correlation hidden between positive and negative feedback. 

Within each session, $\mathbold{E^i}=\{ \mathbold{e_1^i}, \mathbold{e_2^i}, ..., \mathbold{e_n^i}\} \in \mathbb{R}^{n\times d}$ represents the sequence embedding in which the user has interacted with in the $i$-th session. We select the top $n$ most frequent interacted items as input. The value of $n$ depends on different data distributions. 
$d$ is embedding dimension, $\mathbold{F^i}=\{f_1^i, f_2^i, ..., f_n^i\} \in \mathbb{R}^n $ corresponds to frequency information of $E^i$. Here we enhance the confidence of interactions by multiplying each behavior sequence embedding by the corresponding frequency value: 
\begin{equation}
\begin{aligned}
   \mathbold{\hat{e}_j^i} &= f_j^i \ast \mathbold{e_j^i} , j \in [1, n], i \in [1, T]
\end{aligned}
\end{equation}
where $\ast$ is scalar-vector product. Meanwhile, inspired by local activation unit~\cite{zhou2018deep} which adaptively calculate the activation stage of each behavior
embedding, we develop a distance-based attention mechanism to model the relevance of each positive feedback towards compact negative feedback. Here, we first take a frequency weighted sum pooling on negative feedback to generate a fixed-length compact negative feedback representation $\mathbold{\hat{e}_{neg}^i}$ to reduce the noises within single negative feedback, since these negative feedback may be caused by various reasons besides dislike~\cite{he2016fast}.
 
\begin{equation}
\begin{aligned}
    &\mathbold{\hat{e}_{j_{pos}}^i} = f_{j_{pos}}^i \ast \mathbold{e_{j_{pos}}^i}, j \in [1, n], i \in [1, T],\\
    &\mathbold{\hat{e}_{neg}^i} = pooling\left(f_{1_{neg}}^i \ast \mathbold{e_{1_{neg}}^i},..., f_{n_{neg}}^i \ast \mathbold{e_{n_{neg}}^i}\right),\\
    &d_j^i = sim\left(\mathbold{\hat{e}_{j_{pos}}^i}, \mathbold{\hat{e}_{neg}^i}\right)\\
\end{aligned}
\end{equation}
$sim\left(\mathbold{\hat{e}_{j_{pos}}^i}, \mathbold{\hat{e}_{neg}^i}\right)$ is the euclidean distance between the $j$-th positive feedback and compact negative feedback in the $i$-th session. Different from the inner-product based attention mechanism, we manage to assign more weights to that positive feedback which is less similar to negative feedback. Then we apply the softmax function to get a normalized attentive weight $\alpha_j^i$ and multiply it with the frequency weighted positive embedding:
\begin{equation}
\begin{aligned}
    &\alpha_j^i = \frac{\exp \left(d_j^i\right)}{\sum_{k=1}^{n} \exp \left(d_k^i\right)}, \\
    &\mathbold{\tilde{e}_{j_{pos}}^i} = \alpha_j^i \ast \mathbold{\hat{e}_{j_{pos}}^i} \\
\end{aligned}
\end{equation}

To get an enhanced representation, we utilize
GRU~\cite{chung2014empirical} to obtain the sequence embeddings. We feed positive feedback to the GRU unit in descending order of frequency instead of temporal order, the gating mechanism within GRU captures the relationship between the high-frequency feedback and the low-frequency feedback in frequency order. Besides, compared with other recurrent models like RNN or LSTM, GRU is computationally more efficient.
Next, we flatten each hidden state of positive feedback to get a fixed-length embedding vector and concatenate with negative embedding together to obtain personalized representation $\mathbold{{p}_x^i}$ in each session:
\begin{equation}
\begin{aligned}
    \mathbold{{p}_x^i}&= concat\left(GRU(\mathbold{\tilde{e}_{j_{pos}}^i}| j=1,2,...n); \mathbold{\hat{e}_{neg}^i}\right)\\
\end{aligned}
\end{equation}

For each session, we follow the same scheme and compute concurrently. Then we split into two computing branches. On one branch, $\mathbold{{p}_x^i}$ is fed into the self-attention unit~\cite{shaw2018self} to get aggregated personalized representation~$\mathbold{{p}_z^i}$ across all sessions, $\mathbold{{p}_z^i}$ is computed as the weighted sum of linearly transformed input elements, weight coefficient $\alpha_t^i$ is computed using
a softmax function:
\begin{equation}
\begin{aligned}
    &\mathbold{p_z^i} = \sum_{t=1}^{T}\alpha_t^i ( \mathbold{W^i} \mathbold{p_x^i}), \\
    &\alpha_t^i = \frac{exp(w_t^i)}{\sum_{t=1}^{T}exp(w_t^i)} \\
\end{aligned}
\end{equation}
where $w_t^i$ is computed using a compatibility function that compares two input elements $\mathbold{{p}_x^i}$ and $\mathbold{{p}_z^i}$ correspondingly. On the other branch, we feed the two UBP embedding vectors~$\mathbold{{p}_x^i}$ and~$\mathbold{{p}_z^i}$ with a progressive relationship into the UBC module to help to construct latent user groups in each session. 

\begin{table*}[t]
  \caption{Statistics of datasets}
  \small
  \label{tab:statis}
  \begin{tabular}{cccccccc}
    \toprule
    Dataset& Users& Items& Samples& tailed user ratio& body user ratio& head user ratio\\
    \midrule
    Musical Instruments & 51,253 & 14,194 & 129,867& 47\% & 34\% & 19\% \\
    Electronics & 622,308 & 70,323 & 1,589,018& 43\% & 35\% & 22\% \\
    Industrial~(sampling data) & 3 million & 4 million & 0.1 billion& 55\% & 19\% & 26\% \\
  \bottomrule
\end{tabular}
\end{table*}
 
\subsection{User Behavior Clustering~(UBC)}
User Behavior Clustering~(UBC) is another module we designed to further release sparsity, where the individual preference for an unseen item can be referred from the users within the same latent group who have interacted with the item. Since UBC can models the interest of a group of people, UBC can be seen as a coarse-grained semi-personalized interest modeling module.

Auto-encoder~(AE) is one special category of Deep learning methods that compress the data into a dense code and then map the code into the reconstruction of the original input. The appeal of AE lies in the fact that they can learn representations in a fully unsupervised way. However, empirical experience tells that learning group embeddings solely from un-preprocessed input features by AE may not promise robust performance~\cite{ma2019dbrec}. We deal with this dilemma very tactfully by the two progressive representations learned from UBP as input and  constructing a heuristic self-supervised AE to enhance the clustering. 

Refer to Figure~\ref{fig:approach}(b), we learn user group embeddings $\mathbold{G_u^i} \in \mathbb{R}^{k \times d_g}$ within each session, group number $k$ is a rather smaller number compared to the number of users, $d_g$ denotes the dimension of group embedding. 
Recall the intermediate representation $\mathbold{{p}_{x}^i}$ and self attention unit output $\mathbold{{p}_{z}^i}$ in UBP, two representations share same feature dimension, meanwhile $\mathbold{{p}_{z}^i}$ is a higher order representation compared to $\mathbold{{p}_{x}^i}$, thus certain progressive relationship hold. 

In UBC, we initialize AE with random weights, then make it learn effective group embeddings from scratch. In the intermediate layer, we set the representation dimension to be $k$, which is exactly equal to the number of user groups, each value of the vector denotes the probability of user belonging to each group, the learned reconstruction representation $\mathbold{\hat{p}_x^i}$ after AE compute as follows:
\begin{equation}
\begin{aligned}
    & \mathbold{\beta^i} =
    softmax\left(\mathbold{W_c^i} \mathbold{{p}_x^i} + \mathbold{b_c^i}\right),\\
    & \mathbold{\mu^i} =
    \sum_{s=1}^{k}\beta_s^i \mathbold{g_s^i},\\
    &\mathbold{\hat{p}_x^i} =
    \sigma\left(\mathbold{W_r^i} \mathbold{\mu^i} + \mathbold{b_r^i}\right)\\
\end{aligned}
\end{equation}
where $\sigma$ is the sigmoid activation function, $\mathbold{W_c^i} \in \mathbb{R}^{k \times d}$, $\mathbold{W_r^i} \in \mathbb{R}^{d \times d_g}$ ,  $\mathbold{b_c^i} \in \mathbb{R}^k $, $\mathbold{b_r^i} \in \mathbb{R}^d$.  $\beta_s^i$ is the $s$-th dimension of vector $\mathbold{\beta^i}$, $\mathbold{g_s^i}$ is the $s$-th row of group embedding matrix $\mathbold{G_u^i}$. Reconstructed user representation $\mathbold{\hat{p}_x^i}$ will be used for the follow-mentioned reconstruction loss in the UBC. 
The learned semi-personalized representation $\mathbold{c_z^i}$ as follows: 
\begin{equation}
\begin{aligned}
    & \mathbold{c_z^i} = \mathbold{g_j^i},\\
     & j =
     \arg\max _s(\beta_s^i), s \in [1, k] 
\end{aligned}
\end{equation}
where $j$ is the label of user group that user has the maximum activation, $\mathbold{g_j^i}$ is the corresponding group embedding in $\mathbold{G_u^i}$.

The intuition behind self-supervision is that users with similar personal preferences are more likely to belong to the same group. So in the learning process, the learned reconstruction representation $\mathbold{\hat{p}_x^i}$ from $\mathbold{{p}_{x}^i}$ after group embeddings projection should close to the high-level personalized representation $\mathbold{{p}_{z}^i}$ to keep the consistency of individual interest learning and group interest learning. 

In loss construction stage, we deploy the contrastive max-margin objective function that is commonly used in previous work~\cite{iyyer2016feuding,socher2014grounded} to minimize the distance between $\mathbold{\hat{p}_x^i}$ and $\mathbold{{p}_{z}^i}$:
\begin{equation}
\begin{aligned}
     \mathcal{L}_g &= \sum_{i=1}^{T}\sum_{j=1}^{p}\max \left(0, 1 - \mathbold{\hat{p}_x^i} \mathbold{{p}_{z}^i} + \mathbold{\hat{p}_x^i} \mathbold{{\hat{p}_x^{i,j}}}\right)\\
\end{aligned}
\end{equation}
Where $\mathcal{L}_g$ is defined as hinge loss that maximize the cosine similarity between $\mathbold{\hat{p}_x^i}$ and $\mathbold{{p}_{z}^i}$ and simultaneously minimize that between $\mathbold{\hat{p}_x^i}$ and negative samples, \{$\mathbold{\hat{p}_x^{i,j}}| j=1,2,3,...p\}$ represent negative embeddings where we randomly select $p$ users as negative users.

\subsection{Prediction Making Layers}
Up to now, we obtain user behavior feature embeddings including individual-based $\mathbold{{p}_{z}}$ and group-based $\mathbold{c_{z}}$ which concatenate each session output respectively:
\begin{equation}
\begin{aligned}
    \mathbold{{p}_{z}}&= concat\left(\mathbold{{p}_{z}^1}; \mathbold{{p}_{z}^2};...;\mathbold{{p}_{z}^T}\right),\\
    \mathbold{c_{z}}&= concat\left(\mathbold{c_{z}^1};\mathbold{c_{z}^2};...;\mathbold{c_{z}^T}\right)\\
\end{aligned}
\end{equation}
To model the mutual information between the target item and user behavior, we apply the target item attention mechanism to automatically learn the weight of $\mathbold{{p}_{z}}$ and $\mathbold{c_{z}}$. The target item's embedding $\mathbold{e_t}$ contains the information of its ID, price, brand, shop, category. All the vectors are concatenated together to obtain the overall representation vector for the target item. 
Here, we apply the common-used dot-product attention mechanism:
\begin{equation}
\begin{aligned}
    \mathbold{e_{p}} &= \frac{\exp\left(\mathbold{{p}_{z}} \mathbold{W_p} \mathbold{e_t}\right)}{\exp\left(\mathbold{{p}_{z}} \mathbold{W_p}  \mathbold{e_t}\right) + \exp\left(\mathbold{{c}_{z}} \mathbold{W_c} \mathbold{e_t}\right)} \ast \mathbold{{p}_{z}}, \\ 
    \mathbold{e_{c}} &= \frac{\exp\left(\mathbold{c_{z}}  \mathbold{W_c}  \mathbold{e_t}\right)}{\exp\left(\mathbold{{p}_{z}}  \mathbold{W_p}  \mathbold{e_t}\right) + \exp\left(\mathbold{{c}_{z}}  \mathbold{W_c}  \mathbold{e_t}\right)} \ast \mathbold{{c}_{z}}\\ 
\end{aligned}
\end{equation}
where $\mathbold{W_p} \in \mathbb{R}^{d_p \times d_t}$, $\mathbold{W_c} \in \mathbb{R}^{d_c \times d_t}$, $d_p$ is the dimension of $\mathbold{{p}_{z}}$, $d_c$ is the dimension of $\mathbold{c_{z}}$ and $d_t$ is the dimension of $\mathbold{e_t}$. The attention-weighted user behavior feature, target item component feature together with other context and cross features are concatenated together and fed into a multilayer perceptron~(MLP) to get two probabilistic logits, which indicate the probability of this sample belonging to the negative or positive class, respectively.

The learning module is updated by cross-entropy loss, where $\mathcal{D}$ is the training set, $y$ is the label, e.g., for CTR prediction task, $y\in\{0,1\}$ which represents whether the user clicks target item or not, $\hat y_{uv}$ is the predicted score given from our model, $x$ is the input feature and $N$ is the sample count.
\begin{equation}
\begin{aligned}
    \mathcal{L}_{c} &=
    -\frac{1}{N} \sum_{(x,y)\in \mathcal{D}}\left[y \log \hat{y}_{uv} + \left(1-y\right) \log\left(1-\hat{y}_{uv}\right)\right]\\
\end{aligned}
\end{equation}

In the proposed HIM model, we define two loss functions, including group loss $\mathcal{L}_g$ and cross entropy loss $\mathcal{L}_c$. The joint loss can be represented as:
\begin{equation}
\begin{aligned}
    \mathcal{L} &=
    \alpha\mathcal{L}_g+\mathcal{L}_c\\
\end{aligned}
\end{equation}

Where $\alpha$ controls the trade-off between the loss of personalized interest modeling and semi-personalized interest modeling. The training of HIM can be decomposed into two parts since user embeddings are shared for learning group-item and user-item interactions, user embeddings can be well learned from the group-item interactions even the user have few user-item interactions.

\section{Experiments}
\label{sec:experiment}

\subsection{Experimental Settings}
To thoroughly evaluate the performance of HIM, we conduct comprehensive experiments on two public datasets and one real-world dataset. 

\textbf{Public Dataset}\hspace{2mm}
In this paper, we select a widely used public benchmark -- the Amazon dataset~\cite{ni2019justifying}, conducting contrast experiments on two subsets: Amazon's Musical Instruments and Electronics, which contain necessary product reviews and metadata. 
To keep consistent with our feedback setting, we additionally label the data samples whose ratings are higher than 3 as positive 
label while ratings of 1, 2, and 3 are negative~(ratings range from 1 to 5). For each positive feedback, we randomly sample other 5 items as negative feedback. To focus on the recommendation for the long-tailed user, we only filter out users with fewer than 1 positive items and the items with fewer than 5 users, leaving more long-tailed users in our dataset compared with processed datasets used in previous studies~\cite{zhou2019deep,ma2019dbrec}. 
We divide 70\% as training set, 10\% as validation and 20\% as testing.

\textbf{Industrial Dataset}\hspace{2mm} To the best of our knowledge, no public datasets with complete long-tailed user interactions have been released. Public dataset like Amazon usually lacks real negative feedback. So
we also experiment on the industrial dataset, which is a sampling version of the whole data. The data is constructed by impression
and click logs from Lazada online recommender system. Lazada as a growing company, set up an e-commerce business across multiple Southeast Asian countries. In experiment practice,
train, validation, and test set are split along the time sequence, which is a traditional industrial setting. Table~\ref{tab:statis} lists the statistics. 

\begin{figure}[t]
\vspace{-0.5cm}
\subfigure[]{
\label{Fig.sub.session_public}
\includegraphics[width=0.23\textwidth]{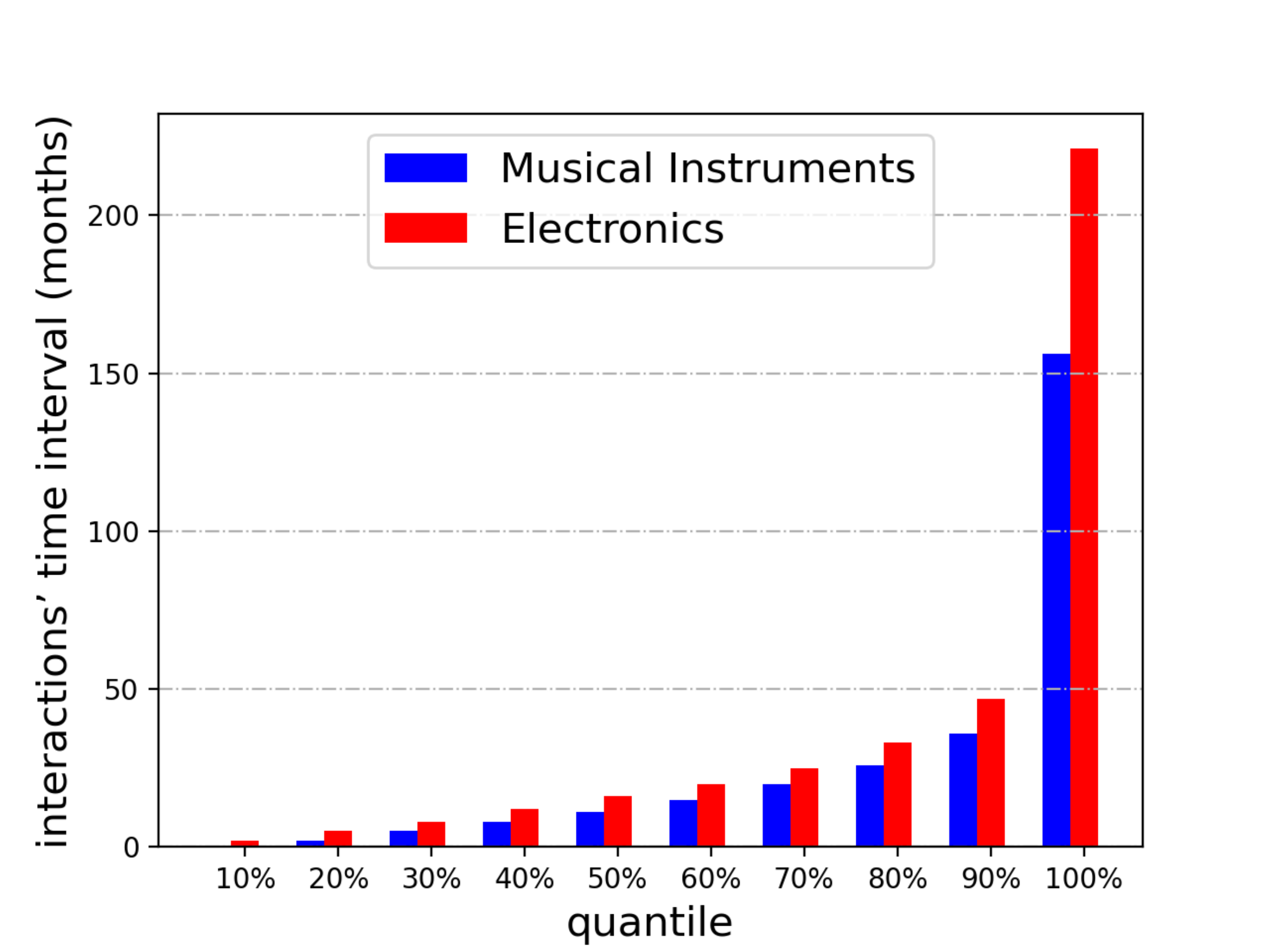}}
\subfigure[]{
\label{Fig.sub.session_lazada}
\includegraphics[width=0.23\textwidth]{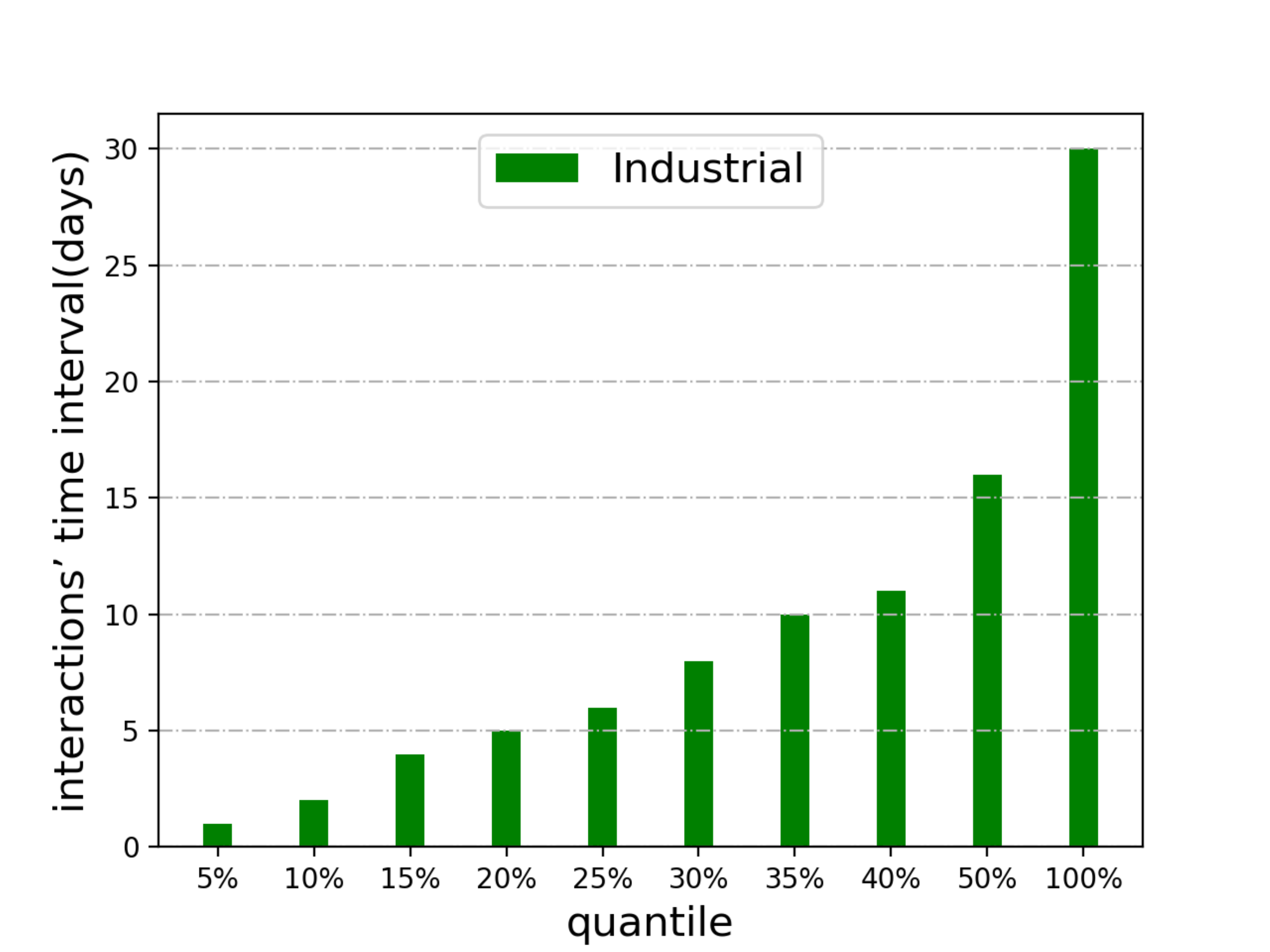}}
\caption{The distribution of user interactions time interval in Public and Industrial datasets, respectively.}
\label{fig:timewindow}
\end{figure}

\textbf{Experimental Setup}\hspace{2mm} In HIM, we subdivide interactions into different sessions. Figure~\ref{fig:timewindow} shows the distribution of user interactions' time interval in Public and Industrial datasets, respectively. We choose the time interval corresponding to approximately  10\%, 30\%, 50\% distribution. For Musical Instruments, session list is \{$14d$, $6m$, $12m$, $all$\}, which contain users' interactions during the last 14 days, interactions during the last 6 months to the last 14 days, interactions during the last 12 months to the last 6 months and earlier interactions, respectively. For Electronics dataset, session list is \{$3m$, $9m$, $18m$, $all$\}. For Industrial dataset, we select the user’s interactions in the last month, session list is \{$3d$, $7d$, $14d$, $30d$\}. 
As previously stated, the user behavior pattern varies greatly. It's inaccurate to report all users' results uniformly when we highlight effectiveness on long-tailed users who have fewer interactions. Hence we divide users into three categories based on users'  interaction frequencies. 
For public datasets, tailed user means users’ behavior sequence length is shorter than 3; the behavior sequence of body user is between 3 and 5; the behavior sequence of head user is more than 5. For the industrial dataset, tailed user means users occurred in the app less than 7 days in one month; the body user is between 7 and 15 days; the head user is more than 15 days. 
As for performance metrics, the area under the ROC curve~(AUC) is adopted, all experiments are repeated 10 times and the averaged result is finally reported.

\textbf{Implementation}\hspace{2mm} HIM is implemented by TensorFlow and we use the Adam~\cite{kingma2014adam} as the optimizer. All related experiments are conducted on one Tesla P100 GPU with 16 GB memory. In the public dataset, we set the batch size to be 256, the learning rate to be $1e-3$, the preliminary embedding dimension is 8, and the group expression dimension is 16. As for the industrial dataset, the batch size is 1024; learning rate is $1e-4$; embedding and the group embedding dimension is 32 and 128, respectively; the dimension setting of MLP layers as [256, 128, 32, 2].

\begin{table*}[t]
  \caption{Results of Different Model(AUC)}
  \small
  \label{tab:results-m}
  \setlength{\tabcolsep}{2mm}{
  \begin{tabular}{p{2cm}<{\centering}p{2cm}<{\centering}|p{2cm}<{\centering}p{2cm}<{\centering}p{2cm}<{\centering}p{2cm}<{\centering}}
    \toprule
    Datasets& Model& all & tailed user&  body user& head user \\
    \midrule
    \multirow{7}*{Musical} & Popularity& 0.798 &  0.796& 0.800& 0.807 \\
     & LR & 0.813 & 0.810& 0.814& 0.823 \\
     & BaseModel & 0.817 & 0.809& 0.820& 0.835\\
     & GRU4Rec& 0.817 & 0.812& 0.818& 0.830 \\
     & DIEN& 0.820& 0.813& 0.826& 0.836 \\
     & DBRec& 0.827& 0.823& 0.825& 0.840 \\
     & \textbf{HIM}& \textbf{0.831}& \textbf{0.825}& \textbf{0.833}& \textbf{0.848}\\ 
    \hline 
    \multirow{7}*{Electronics} & Popularity& 0.811 & 0.811& 0.812& 0.813 \\
     & LR & 0.867 & 0.866& 0.869& 0.869 \\
     & BaseModel & 0.874 & 0.870& 0.878& 0.888 \\
     & GRU4Rec& 0.876 & 0.873& 0.879& 0.886 \\
     & DIEN& 0.878 & 0.875& 0.882& 0.890 \\
     & DBRec& 0.874 & 0.871& 0.877& 0.884\\
     & \textbf{HIM}& \textbf{0.883}& \textbf{0.880}& \textbf{0.887}& \textbf{0.896}\\ 
     \hline 
    \multirow{7}*{Industrial} & Popularity& 0.555 & 0.566& 0.556& 0.549 \\
     & LR & 0.555 & 0.553& 0.555& 0.557\\
     & BaseModel & 0.573 & 0.557& 0.576& 0.580\\
     & GRU4Rec& 0.576 & 0.556& 0.579& 0.583\\
     & DIEN& 0.579 & 0.550& 0.585& 0.585\\
     & DBRec& 0.574 & 0.560& 0.574& 0.578 \\
     & \textbf{HIM}& \textbf{0.607 }& \textbf{0.610}& \textbf{0.612}& \textbf{0.608} \\ 
  \bottomrule
\end{tabular}}
\end{table*}

\subsection{Compared Methods}
In the experiment, we compare HIM with the following methods:
\begin{itemize}
\item \textbf{Popularity}~\cite{cremonesi2010performance} is an intuitive and very simple base method that recommends items by their popularity which is measured by the number of interactions.
\item \textbf{LR~(Logistic Regression)}~\cite{mcmahan2013ad} is a widely used shallow model before deep learning networks for recommendations. 
\item \textbf{BaseModel} is the baseline of HIM. It shares the same Embedding\&MLP setting with HIM. BaseModel simply uses sum pooling operation to aggregate behaviors then reports results, while neither considering negative feedback information nor distinguishing individual and group characteristics. 
\item \textbf{GRU4Rec}~\cite{hidasi2015session} is the first work using the recurrent cell GRU to model temporal sequential user behaviors.
\item \textbf{DIEN}~\cite{zhou2019deep} achieves SOTA performance on sequential recommendation. The key points reside in two-part: one is extracting latent temporal interests from explicit user behaviors, the other is modeling interest evolving process. 
\item \textbf{DBRec}~\cite{ma2019dbrec} is a current SOTA solution for long-tailed user recommendation, models user-user group and item-item group hierarchies for learning compact user/item representation.
\end{itemize}

In results reporting, all involved hyperparameters setting of these methods keep the same with the original work. 
\subsection{Results and Discussion}
\textbf{Model Comparison}\hspace{2mm} 
Table~\ref{tab:results-m} shows the results on Public and Industrial datasets respectively. For a fair comparison, the LR model computes in a straight way without overmuch feature engineering work. According to the experimental results, we have the following observations: 
\begin{itemize}
    \item Compared with naive Popularity, LR gains overall performance improvement on all user categories, which indicates the basic fitting model is still capable of capturing some interactions. However, simple deep BaseModel surpasses both Popularity and LR. It also suggests the deep model is necessary here for modeling complex interactions.
    \item Among deep-learning based models, GRU4Rec simply models chronological behavior. The improvement is little compared to the base model since the temporal dependence between behavior sequences is not obvious for long-tailed users here. DIEN performs better than GRU4Rec, especially for head users, since DIEN captures interest evolution from concrete behavior sequences. While GRU4Rec lacks an explicit modeling process of latent interest behind the interactions. DBRec outperforms DIEN in tailed users on the Amazon Musical and Industrial dataset. This is because the long-tailed issue is more severe on these two datasets. The group information can boost recommendation performance when interactions are insufficient. 
    \item Compared with results on public datasets, AUC on industry dataset is relatively lower. This is because the data sparsity problem is more severe on our system and we use complete and real-world long-tailed user interaction data which makes predicting users' interest even more challenging.
    \item Our proposed model HIM significantly outperforms other methods by a large margin, which proves HIM can learn the long-tailed users' behavior characteristics better, capture user interest with limited behaviors more accurately.
\end{itemize}

\begin{table*}[t]
  \caption{Results on Different Components of HIM}
  \label{tab:results-c}
  \small
  \setlength{\tabcolsep}{0.2mm}{
  \begin{tabular}{p{1cm}<{\centering}p{3.5cm}<{\centering}|p{2.5cm}<{\centering}p{1.5cm}<{\centering}p{1.5cm}<{\centering}p{1.5cm}<{\centering}}
    \toprule
    Datasets&Components & all & tailed user & body user& head user\\
    \midrule
    \multirow{3}*{Musical}& BaseModel & 0.817 &0.809&0.820&0.835 \\
    &+ UBP & 0.826 & 0.820&0.827&0.843  \\
    & \textbf{HIM} (+ UBP\&UBC) & \textbf{0.831} & \textbf{0.825}& \textbf{0.833}&\textbf{0.848}\\
    \hline
    \multirow{3}*{Electronics}& BaseModel & 0.874 &0.870&0.878&0.888 \\
    &+ UBP & 0.879 & 0.876&0.882&0.891  \\
    & \textbf{HIM} (+ UBP\&UBC) & \textbf{0.883 } & \textbf{0.880}& \textbf{0.887}& \textbf{0.896} \\
    \hline
    \multirow{3}*{Industrial}& BaseModel & 0.573 &0.557&0.576&0.580 \\
    &+ UBP& 0.604& 0.606&0.609&0.605 \\
    &\textbf{HIM} (+UBP\&UBC) & \textbf{0.607} & \textbf{0.610}&\textbf{0.612}& \textbf{0.608}\\
  \bottomrule
\end{tabular}}
\end{table*}

\textbf{Ablation Study}\hspace{2mm}In this section, we conduct ablation study to further discuss the contribution of each module in HIM. Since there is no authentic negative feedback in the Public datasets, the UBP module learned in the public datasets is a variant version that excludes the modeling of negative feedback, which refers to positive UBP without pooling operation on negative feedback sequence and without distance-based attention operation between positive and negative feedback. Despite this variation of feature organizing, positive UBP still brings a large gain lift on these datasets as shown in  Table~\ref{tab:results-c}, demonstrating that the behavioral characteristics modeling based on frequency rearrangement is useful for long-tailed users. It is worth to mention that, compared to DIEN, which also only uses positive feedback to personalize interest modeling, positive UBP of HIM performs better than DIEN. 

In the Industrial dataset,  the positive and negative feedback modeling results in UBP are shown in  Table~\ref{tab:results-c}. When we add negative feedback pooling operation to positive UBP, it brings 0.015 absolute AUC gain, which indicates the negative feedback is a highly informative supplement to basic behavior. 
Meanwhile, by adding the distance-based attention between positive and negative feedback, which can further model users' behavior and generate a ``reliable'' user representation, it achieves 0.003 absolute AUC gain. 
In addition to personalized modeling module UBP, semi-personalized modeling module UBC also brings extra absolute AUC lift on Public and Industrial datasets, indicating that semi-personalized recommendation is a good supplement for personalized recommendation, especially for those long-tailed users.

\begin{figure}[t]
\vspace{-0.5cm}
\subfigure[]{
\label{Fig.sub.dist}
\includegraphics[width=0.23\textwidth]{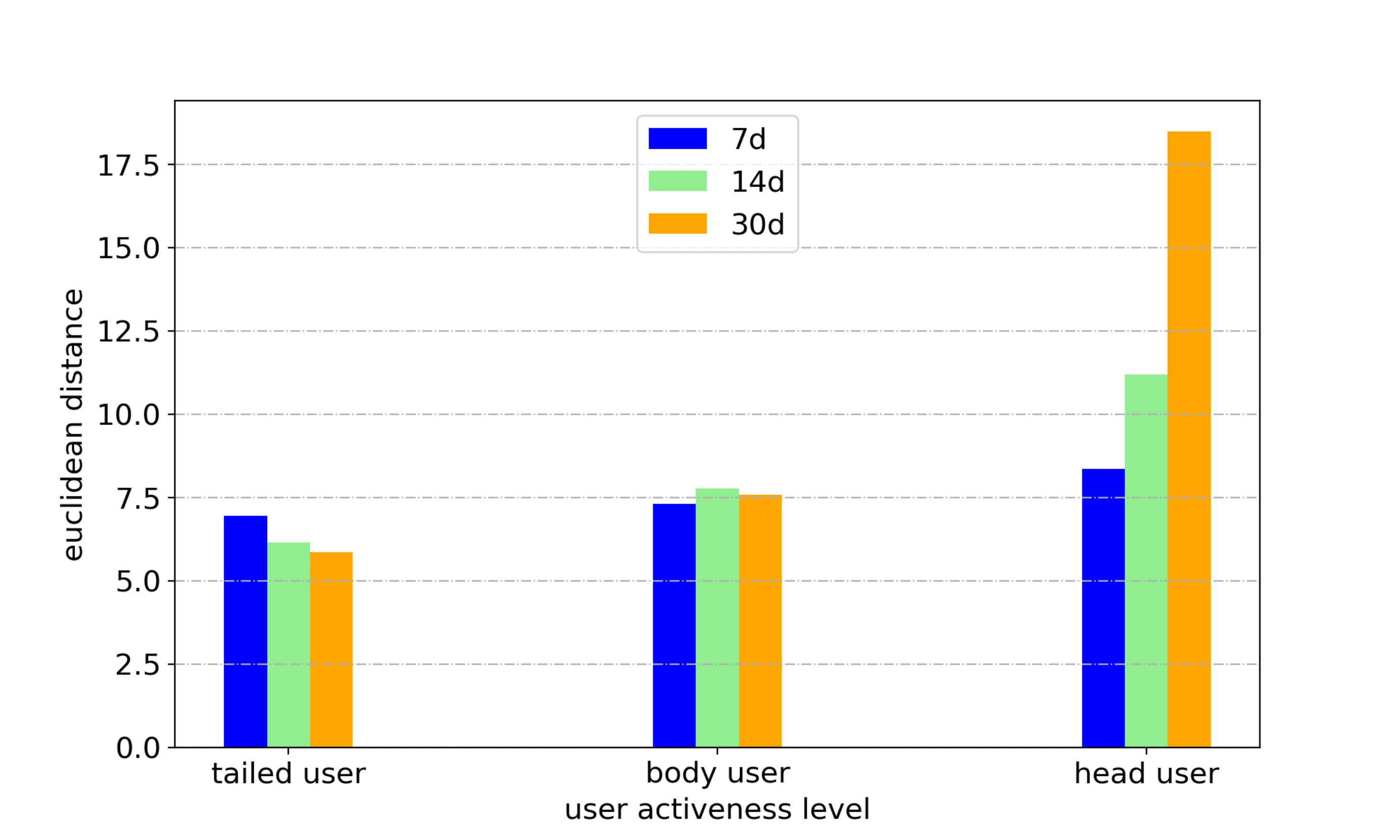}}
\subfigure[]{
\label{Fig.sub.att}
\includegraphics[width=0.23\textwidth]{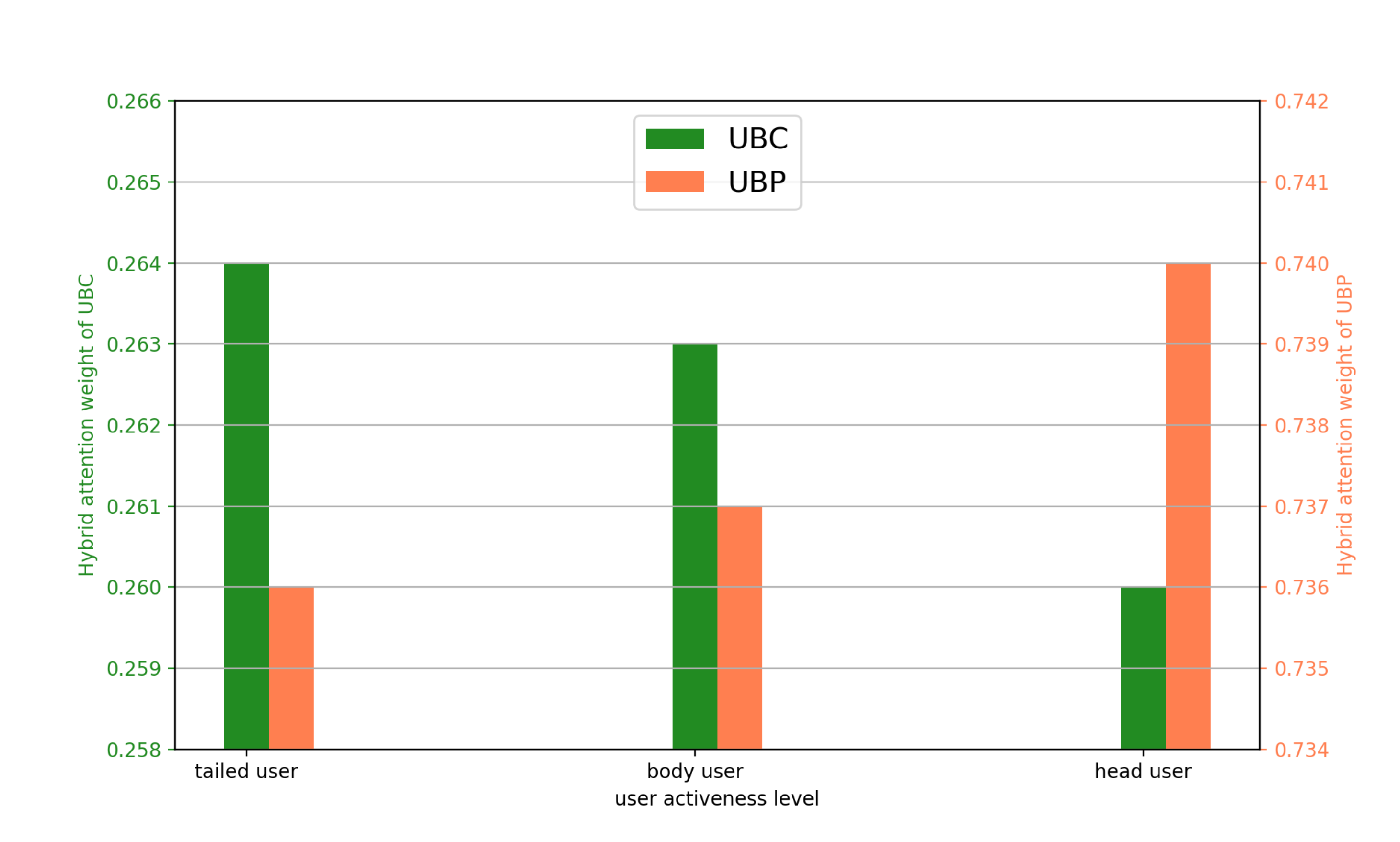}}
\caption{Performance under different interaction sparsity: (a) shows the euclidean distance between positive and negative feedback, the smaller euclidean distance reflects the higher uncertainty of positive behaviors. (b) shows the weights allocation between UBP and UBC, users with higher sparsity have a higher UBC weight.
}
\label{Fig:activenes}
\end{figure}

\textbf{Effect of Confidence Modeling in UBP}\hspace{2mm} We argue that there exists a random click in positive feedback. Thus we design the euclidean distance-based attention to model the similarity between negative aggregated embedding and each positive item embedding. Refer to Figure~\ref{Fig:activenes}(a), we visualize the distance for users of different sparsity. We have the following observations: First, the distance for tailed users is relatively small, the smaller euclidean distance reflects the higher uncertainty of positive behaviors. Second, for head users, distance varies largely among different sessions. This indicates that users' behavior shifts across time, the session can help understand users' preferences instead of modeling the whole behavior sequence directly. 

\textbf{Effect of UBC}\hspace{2mm} When we hybrid the embedding from UBP and UBC, we apply the target item embedding to automatically learn their weights. To study the effect of this attention and have a better understanding of UBC, we analyze the learned weight and drill down to the different types of users. As shown in Figure~\ref{Fig:activenes}(b), UBP's weight is consistently higher~(higher than 0.7) than UBC's weight~(around 0.26), indicating that personalized embedding is more important when inferring users preference. Then, UBC's weight is relatively higher for tailed users. This indicates that such semi-personalized group embedding is a promising method when users' interaction is extremely sparse.

\begin{figure}[t]
\vspace{-0.5cm}
\subfigure[]{
\label{Fig.sub.group_public}
\includegraphics[width=0.2\textwidth]{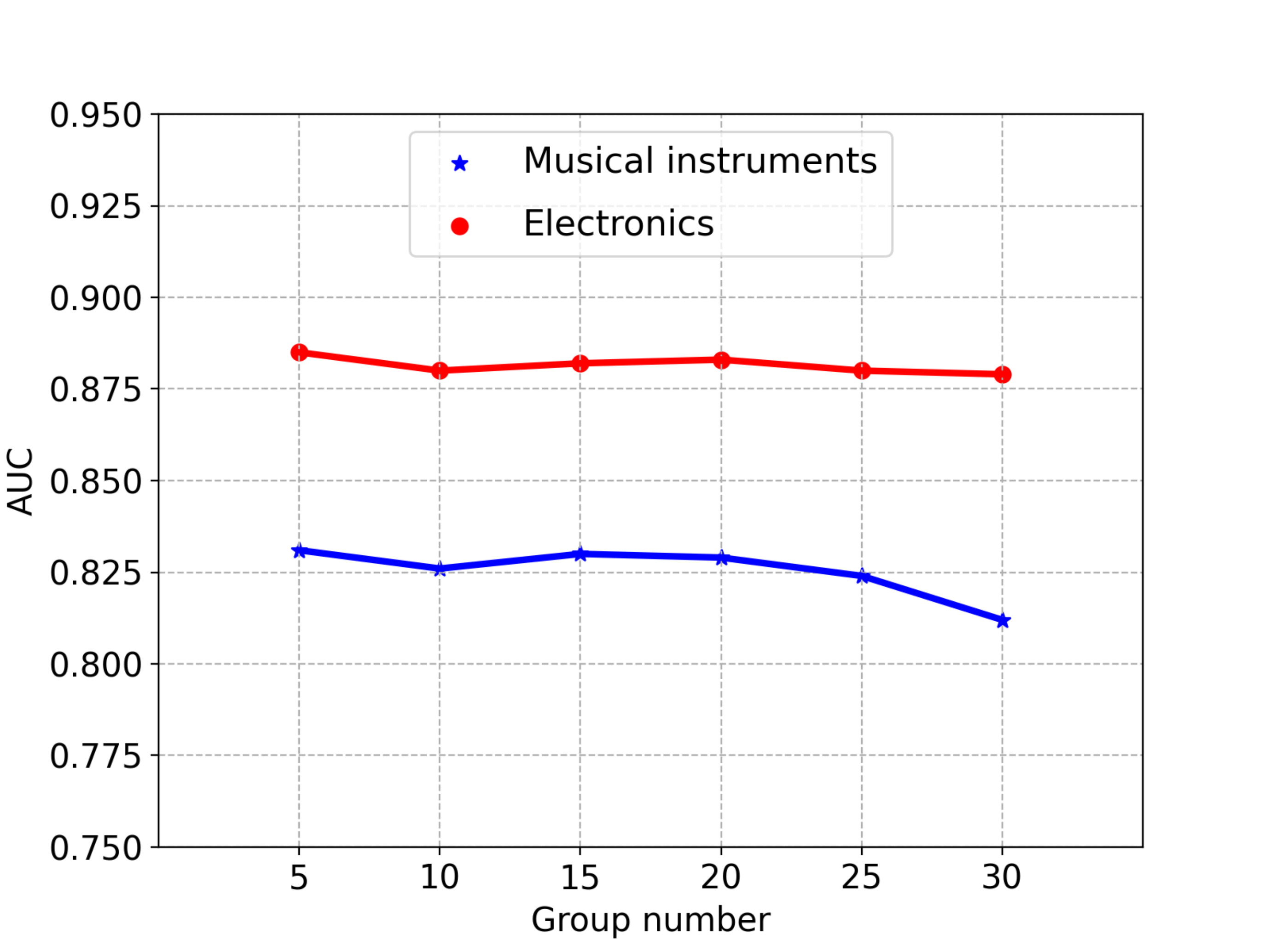}}
\subfigure[]{
\label{Fig.sub.loss_public}
\includegraphics[width=0.2\textwidth]{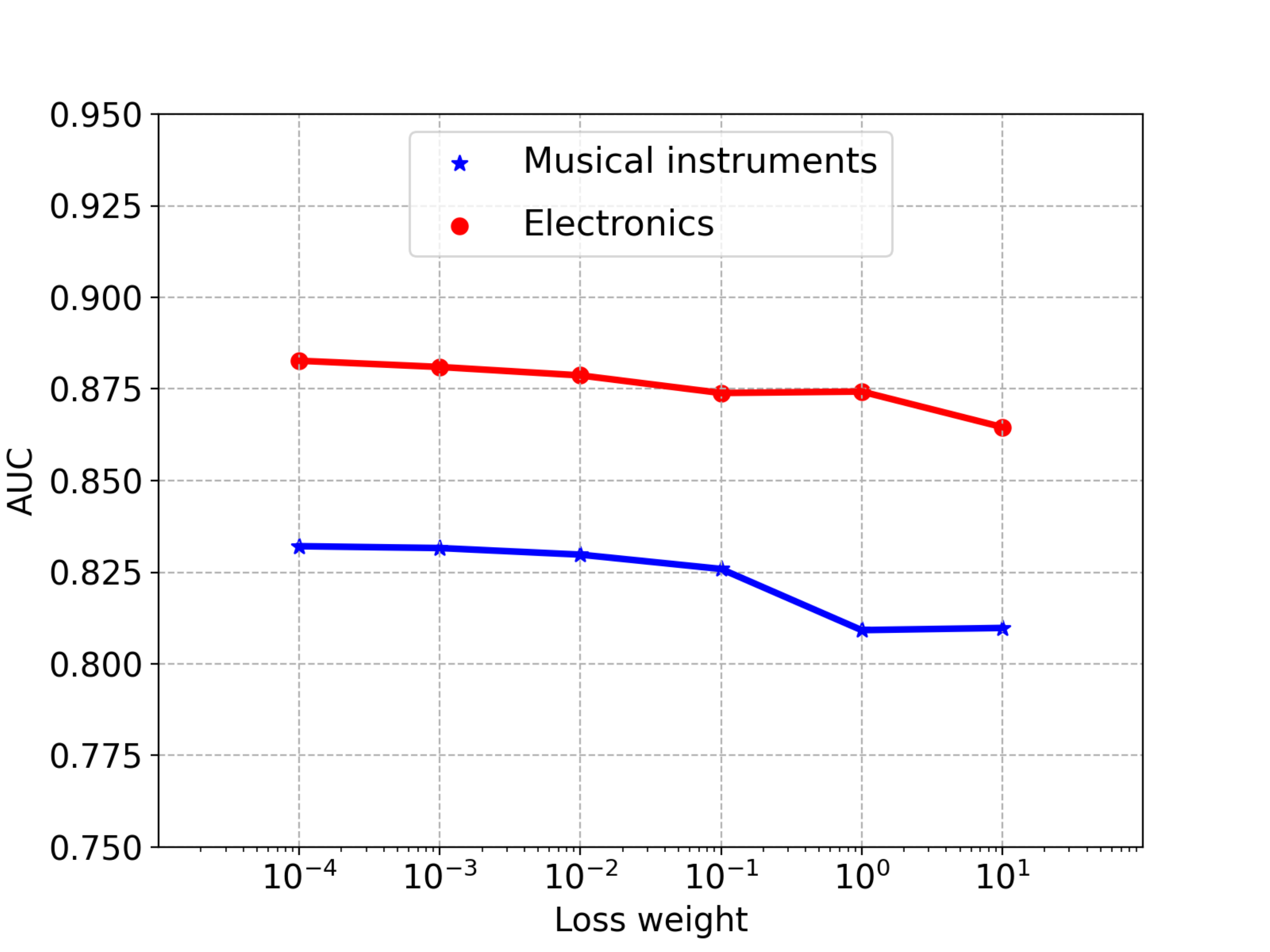}}
\subfigure[]{
\label{Fig.sub.group_lazada}
\includegraphics[width=0.2\textwidth]{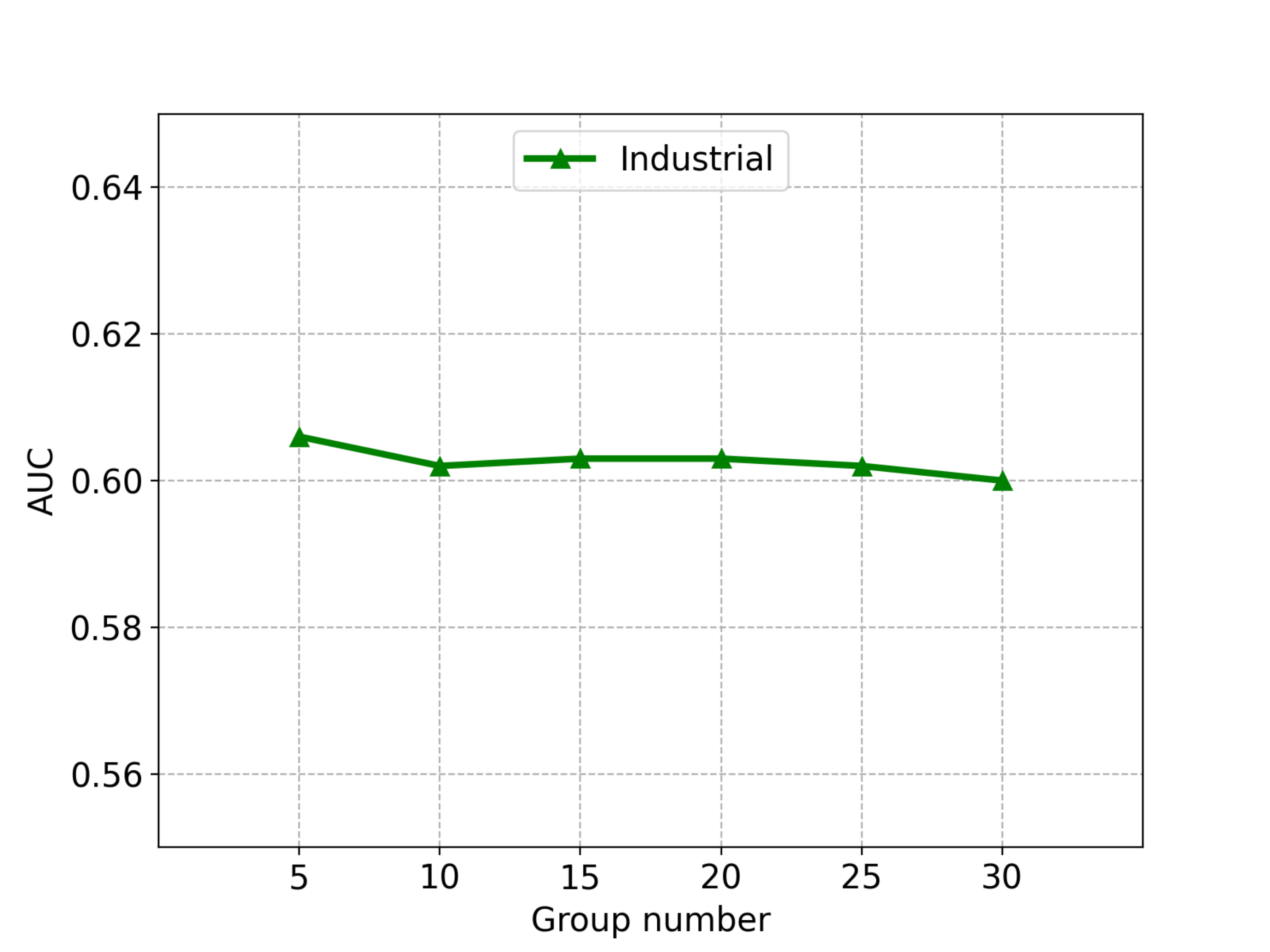}}
\subfigure[]{
\label{Fig.sub.loss_lazada}
\includegraphics[width=0.2\textwidth]{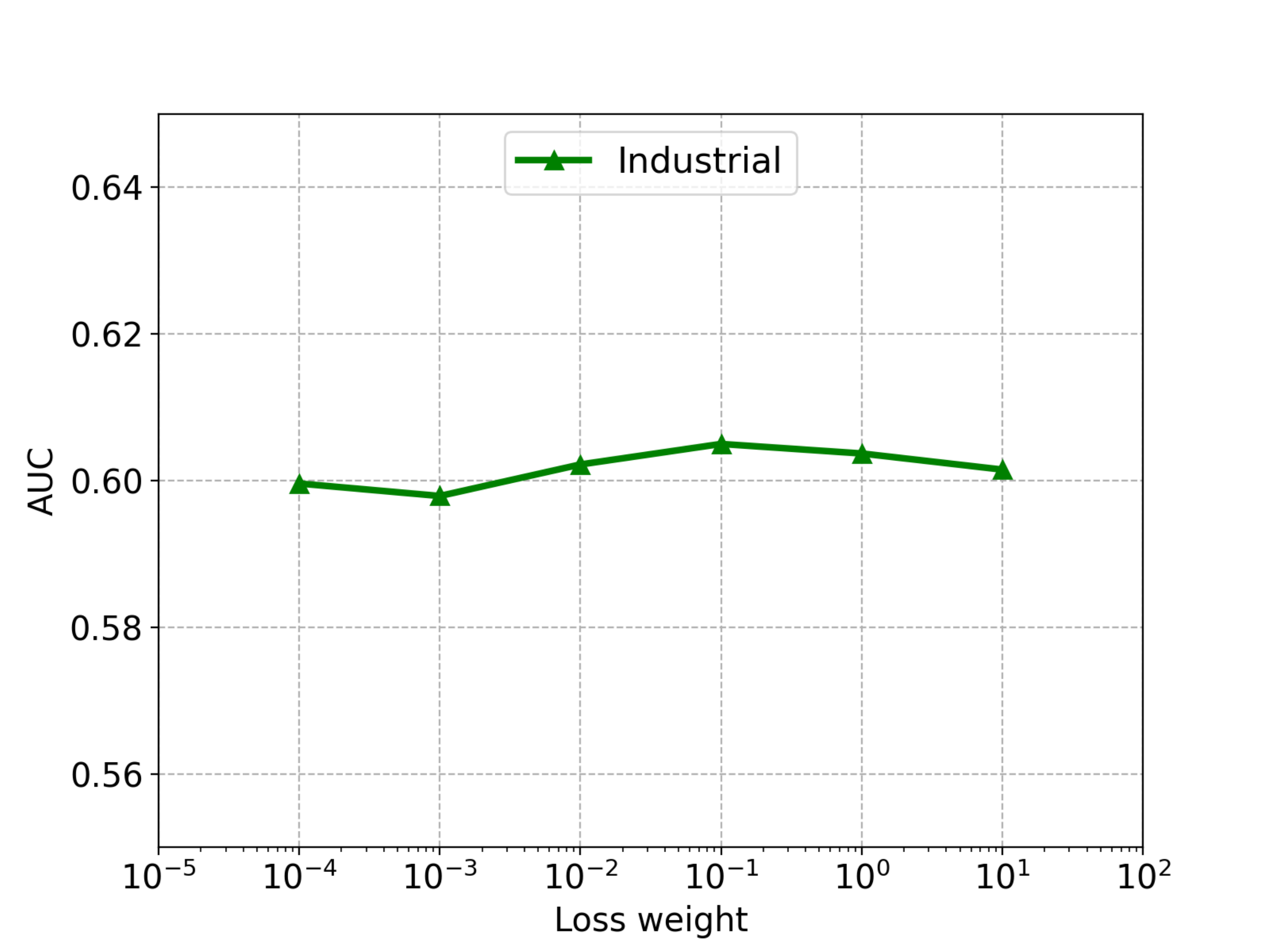}}
\caption{Performance of tuning hyperparameters.}
\label{fig:hyper}
\end{figure}

\textbf{Effect of hyperparameters}\hspace{2mm}
In this section, we further discuss the performance of HIM under different pre-specified user group number $k$, and the loss weight $\alpha$ for hybrid modeling in a limited set. Empirically, group number is varied amongst [5,10,15,20,25,30], and the $\alpha$ [0.0001,0.001,0.001,0.1,1,10]. As shown in Figure~\ref{fig:hyper}, we present the performance across the datasets, Figure~\ref{fig:hyper}(a) and Figure~\ref{fig:hyper}(b) shows the performance has a rapid degradation with group number $k$ = 30 and loss weight $\alpha$ = 1 on Amazon Musical instruments dataset. Overall the performance is relatively stable under a wide-range choice and achieves the best performance when clustering into 5 groups. In public datasets, the best performance achieves with $\alpha$ = 0.0001 while for Industrial dataset is 0.1. Our analysis result is proved to be related
with the size of the dataset.

\subsection{Online A/B testing}
We have deployed the proposed HIM in Lazada online recommender scenario across different Southeast Asian countries, including Indonesia~(ID), Malaysia~(MY), Vietnam~(VN), Thailand~(TH). A standard A/B test is conducted online, we perform the online experiments for one month, and the average item page view (IPV) Gain of different user groups are reported. As shown in Table~\ref{tab:results-o}, all users' IPV are improved. A higher IPV indicates that users are more willing to browse and click items on our platform. Especially for the tailed users, the improvement is large, because HIM can well learn long-tailed users' preference, which leads to more positive feedback. As this e-commerce company is still growing, the optimization based on the characteristics of long-tailed users would result in a significant boost in revenue for the long term.

\begin{table}[t]
  \caption{Online A/B testing Results  on Different Southeast Asian countries}
  \label{tab:results-o}
  \small
  \begin{tabular}{cccccccl}
    \toprule
    Country & all IPV Gain& tailed user& body user& head user& \\
    \midrule
    ID & +7.2\% & +6.7\% & +7.5\%& +7.2\%& \\
    MY & +8.6\% & +11.1\% & +9.6\% & +8.1\%& \\
    TH & +7.5\% & +7.3\%& +8.1\%& +7.5\%& \\
    VN & +10.5\% & +12.5\%& +12.7\%& +9.8\%& \\
  \bottomrule
\end{tabular}
\end{table}

\section{Conclusion}
\label{sec:conclusion}
In this paper, we study user interest modeling under limited interactions in newborn e-commerce, provide a new perspective to consider interaction sparsity issue caused by dominant long-tailed users in an actual production environment. We propose a new structure for long-tailed users, namely Hybrid interest modeling~(HIM) network to balance individual expression and group expression to achieve better recommendation performance with limited behavior data.

In short, HIM is a successful practice for our online campaign and can be an instructive recommendation solution for other similar newborn e-commerce business.

\bibliographystyle{ACM-Reference-Format}
\bibliography{sample-base}

\appendix

\end{document}